\documentclass[11pt,twoside,onecolumn]{article}
\usepackage[]{latexsym,amsmath,amssymb}
\usepackage[]{epsfig}
\newcommand{\be}{\begin{eqnarray}}
\newcommand{\ee}{\end{eqnarray}}

\title{\bf Reply to
\\
``Holographic principle in spacetimes with extra spatial dimensions''
\\
and
\\
``About quantum fluctuations and holographic principle in $(4+n)$-dimensional spacetime''}
\author{Fabio~Scardigli$^a$\thanks{Corresponding author address:
Via~Europa~20, 20097 S.~Donato, Milano, Italy.
E-mail: scardigli@fisica.ist.utl.pt}
$\ $and
Roberto~Casadio$^b$\thanks{E-mail: casadio@bo.infn.it}
\\
\\
{\em $^a$CENTRA, Departamento de Fisica, Instituto Superior Tecnico}\\
{\em Av. Rovisco Pais 1, 1049-001 Lisboa, Portugal}
\\
\null
\\
{\em $^b$Dipartimento di Fisica, Universit\`a di
Bologna and I.N.F.N., Sezione di Bologna,}\\
{\em via Irnerio~46, 40126 Bologna, Italy}}
\date{}
\begin{document}
\maketitle
\begin{abstract}
We reply to the comments by P.~Midodashvili about our previous paper~\cite{SC}.
We argue that, contrary to the conclusions in Refs.~\cite{midoda1, midoda2},
the Generalized Uncertainty Principle proposed by Ng and van~Dam in Ref.~\cite{ngvd}
is compatible with the Holographic Principle in spacetimes with
extra dimensions only for a very special (and somehow unrealistic) choice
of the relation between the size and mass of the clock.
\par
\null
\end{abstract}
\raggedbottom
\setcounter{page}{1}
%
%
%\section{Reply}
%\setcounter{equation}{0}
%
P.~Midodashvili in two recent papers~\cite{midoda1, midoda2} studies
Ng and Van~Dam's Generalized Uncertainty Principle (GUP)~\cite{ngvd}
and concludes that it ``complies with the holographic principle
{\em also\/} in spacetimes {\em with\/} extra-dimensions'', in contrast
with the result previously obtained in Ref.~\cite{SC}.
\par
The arguments of Refs.~\cite{midoda1, midoda2} appear to be
based on the following two hypothesis:
\begin{enumerate}
\item
\label{1}
besides the usual quantum mechanical and gravitational errors
in distance measurements, the size $a$ of the clock must be taken
into account as an additional source of uncertainty in the total error;
\item
\label{2}
the size of the clock can be written as
\be
a=\alpha\,r_{S(4+n)}
\ ,
\label{a}
\ee
where $\alpha$ is a dimensionless coefficient and $r_{S(4+n)}$ the
Schwarzschild radius of the clock in $4+n$~dimensions.
As the clock is not a black hole, one must assume $\alpha>1$.
\end{enumerate}
\par
About the first hypothesis, we note that in Refs.~\cite{midoda1,midoda2}
no physical reason is given to clarify why the {\em whole\/} size $a$ of the
clock should {\em a priori\/} result in a distance error in the measurement
process.
In principle, the clock is a macroscopic object, with a very well defined
geometric shape which is known before the measurement takes place.
Of course, an uncertainty in the numerical value of $a$ can be considered,
but this obviously does not mean the whole size $a$ gives an error in
time measurements.
Indeed, an argument in favour of such an assumption is the following:
one should consider that there is an error $\delta t$ in the registration
time of the clock roughly given by the time a beam of light takes to travel
the clock's size, that is $\delta t\sim a/c$.
This may be considered as the typical accuracy of an ``ideal'' physical
clock.
In general, however, the actual sensor that captures the beam of light
inside the clock will have a size $b$ very likely much smaller than $a$
and the error $\delta t\sim b/c\ll a/c$.
Moreover, this estimate of $\delta t$ only holds if the metric
inside the clock is flat.
If that is the case, one then does not see why it should be related to the
gravitational radius of the {\em whole\/} clock $r_{S(4+n)}$, since there
is no horizon at $r=r_{S(4+n)}$ inside the (flat) clock.
It is clear that very detailed information about the clock inner structure
is needed to estimate the error $\delta t$, for
which it is therefore difficult to come to a general expression.
\par
Bearing the above conclusion in mind, it is now easy to express the
main objection to the second hypothesis of Refs.~\cite{midoda1, midoda2}.
In writing down Eq.~(\ref{a}), the Author assumes that $\alpha$
is a dimensionless number, larger than $1$, but {\em independent}
from the mass $m$ of the clock.
This is tantamount to the size of the clock going like $m^{1/(n+1)}$
for which the holographic scaling in $4+n$ dimensions is then recovered.
We however do not see any reason for using this particular scaling, which
appears as a rather arbitrary claim.
For example, a clock made of a constant density material would instead
have $a\sim m^{1/(3+n)}$ (in $4+n$~dimensions).
To all practical purposes, the size $a$ of a clock cannot be
correlated with its mass $m$ in any universal way and the
hypothesis~\ref{2} therefore sounds too restrictive and, as such, misleading.
%the error stemming from taking $\alpha$ as {\it independent\/} from $m$.
Since in Ref.~\cite{SC} we were looking for an ``ideal'' clock whose
accuracy (including quantum and gravitational errors) is maximum,
the expression of the error in distance measurements was minimized
with respect to the mass of the clock.
How each term in the expression of the error depends on the mass $m$ is
therefore crucial.
\par
From the above observations several conclusions can now be drawn:
\begin{description}
\item[a)]
an additional error $b$ in distance measurements may be allowed, as
indicated in Refs.~\cite{midoda1, midoda2}, of the order of the size $a$
of the clock or, more likely, much smaller ({\em i.e.},~$b\ll a$);
\item[b)]
in any case, the error due to the size of the clock is essentially not
correlated with the mass of the clock, and $b$ must therefore be
considered as a constant (with respect to $m$) in the global error
on the measure of the length $l$ (for the expressions of $l_{\rm QM}$ and
$l_{\rm C}$ see Ref.~\cite{SC}),
\be
\delta l_{\rm tot}(m)=\delta l_{\rm QM}(m)+b+\delta l_{\rm C}(m)
\ ;
\ee
\item[c)]
by extremizing $\delta l_{\rm tot} (m)$ with respect to $m$
in 4~dimensions, one then gets
\be
\frac{\partial l_{\rm tot}(m)}{\partial m}=0
\quad
\Rightarrow
\quad
m_{\rm min}=c\,\left(\frac{\hbar\,l}{G_{\rm N}^2}\right)^{1/3}
\ ,
\ee
where $G_{\rm N}$ is Newton's constant and
\be
\left(\delta l_{\rm tot}\right)_{\rm min}=
2\,\left(\frac{\hbar\,G_{\rm N}}{c^3}\,l\right)^{1/3} +
\left(\frac{\hbar\,G_{\rm N}}{c^3}\,l\right)^{1/3}+b
=3\left(\ell_{\rm p}^2\,l\right)^{1/3} + b
\ ,
\ee
which contains a term proportional to $l^{1/3}$ and $\ell_{\rm p}$ is the
Planck length.
In $4+n$~dimensions we shall have essentially the same result of
Ref.~\cite{SC},
\be
\left(\delta l_{\rm tot}\right)_{\rm min}=N(n)
\left(\frac{\ell_{(4+n)}^{n+2}}{a^n}\,l\right)^{1/3} + b
\ ,
\ee
with $N(n)$ a numerical coefficient and $\ell_{(4+n)}$ the fundamental
gravitational length in $4+n$ dimensions;
\item[d)]
we can write the number of degrees of freedom in a volume of size $l$
in 4~dimensions as
\be
\mathcal{N}(V)=
\left(\frac{l}{\left(\delta l_{\rm tot}\right)_{\rm min}}\right)^3
=
\left(\frac{l}{3\left(\ell_{\rm p}^2\,l\right)^{1/3}+b}
\right)^3
\sim
\left(\frac{l^2}{\ell_{\rm p}^2}\right)
\ ,
\ee
and holography is recovered for $l\gg b^3/\ell_{\rm p}^2$ (ideally, one
could take $b\sim \ell_{\rm p}$ for the minimum size of the sensor).
Instead, in $4+n$ dimensions we have
\be
\mathcal{N}(V) = \left(
\frac{l}{\left(\delta l_{\rm tot}\right)_{\rm min}}\right)^{3+n} \sim
\left(\frac{l}{
\left(\ell_{(4+n)}^{n+2}\,l\, a^{-n}\right)^{1/3} + b}\right)^{3+n}
\ee
and the holographic counting is lost (since $n>0$).
Even for $l\gg a^n\,b^3/\ell_{(4+n)}^{n+2}$ one has
\be
\mathcal{N}(V)
\sim
\left(\frac{l^2 a^n}{\ell_{(4+n)}^{2+n}}\right)^{1+\frac{n}{3}}
\sim \left(\frac{l^2}{\ell_{(4+n)}^2}\right)^{1+\frac{n}{3}}
\ ,
\ee
in which we again considered the rather ``ideal'' case $a\sim\ell_{(4+n)}$.
\end{description}
Hence, the holographic counting is still lost in $4+n$ dimensions, even if we remark here that
the holographic bound is not violated, since $2\,\left(1+{n}/{3}\right) < 2+n$.
\par
To conclude, we believe that the main result in Ref.~\cite{SC}, that holography
does not appear compatible with the GUP of Ref.~\cite{ngvd} in $4+n$~dimensions,
holds in general even
if one properly includes in the computation an error due to the size of the clock.
On the contrary, the result stated in Refs.~\cite{midoda1, midoda2} requires a highly
specific structure for the clock and the corresponding mass scaling~(\ref{a}),
which is clearly not general.
It would be interesting to experimentally test whether such a structure is indeed
``unavoidable'' in our real world or just an artifact to recover the holographic scaling.
\end{document}